Reply to Reply to Comment and Further Analysis of paper
"New probing techniques of radiative shocks, Opt.Comm. **285**, 64 (2012)"

Michel Busquet
*ARTEP,inc – Ellicott City, MD 21042, USA*

**Abstract**

We show in this paper that the "reply to comment" by C.Stehlé et al was not satisfactorily and is contradictory to their latest results. The analysis of the published results of their last campaign reveals an average shock velocity below 20 km/s which should not be compatible with the existence of a radiative precursor. These results also suggest exploding back layers of the pusher and some jet originating from a crater at the back of the pusher.

**1- Introduction**

In C.Stehlé et al, 2011 [1], the authors present results obtained on the PALS facility. They probed with an auxiliary XUV laser at λ=21.2nm, a (presumably radiative) shock launched by a gold coated CH pusher ablated by the infrared PALS beam (at λ=1.315 µm). The recorded image they present in [1] was not convincing and was rather suggesting an "in shot" blurring. The allegded precursor could rather be only the penumbra cast by the pusher edge. The time dependant luminosity recorded by fast diodes was no more convincing once accounting for the corona emissivity. This discussion has been presented in a "comment" [2] which questionned the conclusions they drew, mostly the observation of a radiative precursor.

In their reply to the comment,[3] they still claim to have observed the radiative precursor of an intense shock driven by the accelerated pusher. They show a XUV laser radiography of a naked, empty target, with no laser irradation, which present a good resolution (<10 µm). But this does not preclude a blurred image, overwhelmed by an important speckle pattern, and through a target equipped with Si3N4 windows, filled with Xenon and irradiated by the infrared laser beams.

Furthermore, at the SF2A conference in Montpellier 2013 [4] and at the International conference on black holes, jets and outflow, Kathmandu, 2013, [5] they presents images which definitely exhibit a reproducible 2D structure, thanks to the use of a better mirror. Such a 2D structure cannot be washed out to show a 1D-like image as the one presented in [1] *except by* some "in shot" blurring of a few hundreds microns.

**2- achieved "in shot" resolution**

Once deconvoluted from the speckle pattern, the "in shot" image (Fig.6 top) of Ref.[1] shows a 1D-like structure of the so-called "shock". This image is obtained when firing a target with windows, filled with gas, irratiated by the IR laser creating the "shock", and imaged by the XUV laser @ 21.3 nm. It includes possible blurring from various reasons. The achieved resolution (from geometrical aberrations, refraction+diffraction, gas or plasma blurring, …) is unknown, even if the "static" resolution is better than 15 µm (Fig.2 of Ref. [3]).

With an improved imaging setup, mainly a better XUV mirror, [4,5] but in similar conditions, they obtain a good resolution (which proves the resolution was poor in [1]). The actual structure of the image of the "shock" is 2 D and presents a 150x300 µm oblate bubble, over an exponentially decreasing absorption (with a gradient length of approx. 250 µm, see Sec.3). *Only a blurring of a fraction of the transverse size can give a quasi 1D image of it*, like the one presented in [1]. Then our conclusion in ref.[2] of an achieved "in shot" resolution of 200-400 µm (or rather a 200-400 µm blurring) is validated by their own recent results. The existence of a radiative precursor claimed in [1] is thus still to be demonstrated in these experimental conditions.

Furthermore this bubble shape shadow can hardly be understood as coming from a curved shock but should rather result from diffraction on a depleted jet-like structure. The non perpendical direction of the jet should confirm this possible origin. Obviously a complete 2D simulation of the formation of the bubble is needed for complete analysis of the experiment. Because the drive laser operates at 1.315 µm, hot electrons are created and the back layers of the gold coating may undergo some preheating, which may explain their expansion producing the observed gradient, and the expelled crater as well. This hypothesis again has to be checked with full simulations.

So the pedestal of the darker zone (Fig.1 of this paper and Fig.6 of Ref.[1]) is most probably the edge of the blurred penumbra, or an expansion of the back layers of the preheated pusher, or both. All their conclusions concerning an observation of a radiative precursor have then to be revised.

### 3- analysis of images reported in Refs.[4,5]

As seen from the 3d representation of the XRL image (Figs.5-6), the bubble-like structure is superimposed on an exponentially decreasing attenuation profile, with a gradient length around 250 µm. The absorption plateau correspond to the pusher dense remains. The dark rings are diffraction patterns, and are not the trace of the density peak of the shock as claimed in Refs. [4,5]. The transmission is even larger in the zone the authors called the "shock region" than in the zone they called the "post-shock" region! Both regions lie in the gradient zone, extending 400 µm after the 300 µm long non-transmitting zone. No jump, or peak, of density can be inferred from the recorded XRL image in this gradient region. But the peak of absorption, probably due to a peak in density, appears to be located just at the edge of the black zone, at approx. 0.3 mm from the end (Fig.5), which means an average shock speed & piston velocity of approx. 18 km/s, using the numbers given in ref.[4,5]. The jet-like structure may be due to fast particles, or to some matter ejected from the crater seen in the radiographies. In Ref.[5] this "jet" apparently performs some "snow-plowing" in what could be a density gradient (Fig.6).

The origin of this attenuation gradient, apparently a density gradient, and of the snow-plow feature requires
- the knowledge of the monochromatic opacity variation with temperature and density, though first estimation with STA shows that the variation with temperature is expectedly small, and one have rather to look at the density variations.
- 2D hydro-simulations (or even 3D for an oblique jet) including effects from the suprathermal particles, ablation of the rear surface of the piston, matter ejection from crater, …

Note that analysis of interferograms precedently obtained at PALS leads also to some expansion of the back layers, however negligible when using the third harmonic.[6]

### 4- xenon absorption of the XRL beam

We now come to the possible contrast of absorption between cold Xe and precursor. As the temperature in the precursor (if it exists) is a ramp and not a step, the absorption would present also a ramp from 0% at the front of the precursor to around 15% a fraction of mm away. Contrarily to the affirmation by C.Stehlé et al, "*the corresponding relative variation of 15% [of absorption through the whole width of the tube], spread over several hundreds of microns in the precursor should be detected*", the absorption ramp would not be differenciated from background variations. On the other hand the absorption is proportional to the ion density, then the observed attenuation gradient is probably due to a density gradient.

The origin of it could possibly be from expanding back layers of the pusher. Once again, the allegedly observation of a radiative precursor, a pure temperature wave at constant ion density, is then highly disputable. As ionization increase with temperature, the radiative precursor is also an electron density wave and would be better probed (if it exists) by interferometry.[6]

**In conclusion:**

- no radiative precursor has been evidenced in the experiment reported in Ref.[1] and no more in Ref.[4,5]
- an "in shot" resolution/blurring of 200 to 400 µm is the best interpretation of the 1D-like image presented in Ref.[1]
- the hydro shock (i.e. the density peak) is more probably located at the edge of the absorbing plateau, with an average velocity below 20 km/s
- the density gradient at the back of the pusher has to be explained, possibly by some exploding back layers of the pusher
- crater, jet and diffraction patterns seems to be reproducible features
- x-ray or XUV radiography probes density variations, not temperature variations. *Therefore it will not probe any radiative precursor.*

Note also that the targets used in [1,4,5], with full window walls, are directly derived from the one described in [7], though the design of the target holder is new.

**Figures :**

1- XRL image of a fired target, from Fig.6 of Ref.[1], and transversally averaged attenuation profiles. Superimposed to the speckles, the darker zone is attributed by C.Stehlé to absorption by a radiative precursor.

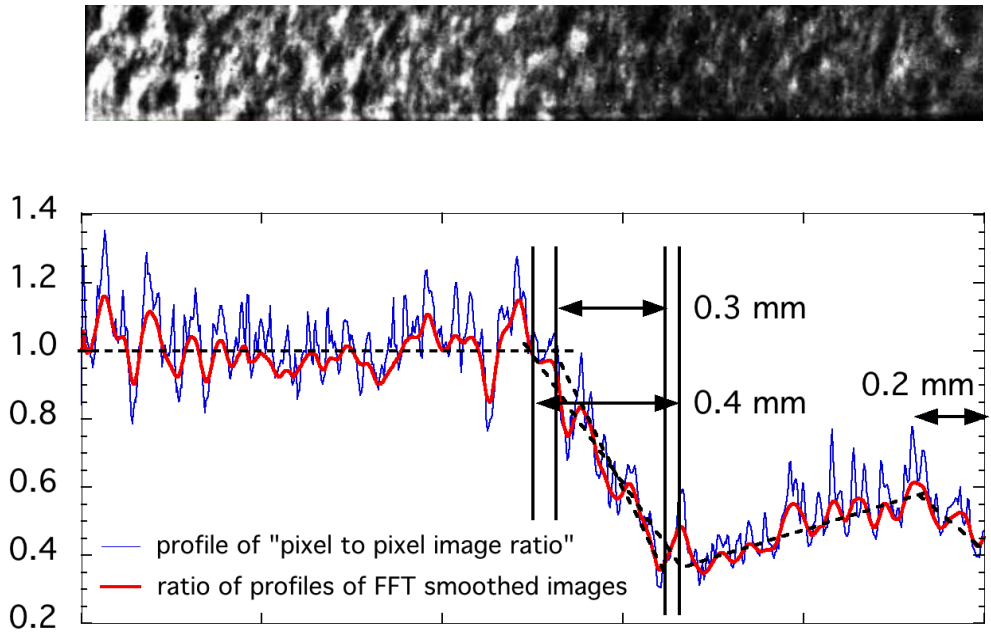

2- XRL image of a naked, void and unfired target, from Fig. 2 of Ref [3]. Note that the speckle patterns differ from Fig.1, contrarily to the pair of image in Fig.6 of [1], which means that the position of the optical components has been changed in between.

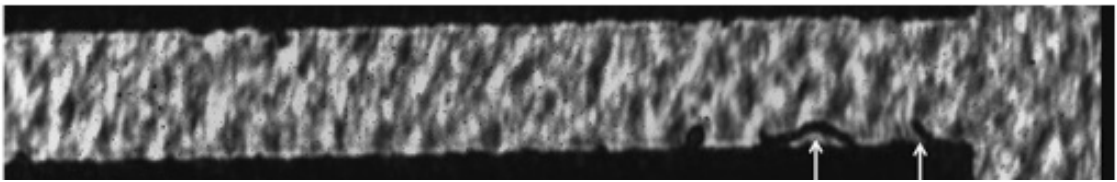

Fig.3- XRL image of a fired target, from [4]. A better XUV mirror has given well resolved images. The shadow cast by the dense pusher exhibit a crater at the origin of the jet-like structure. Diffraction patterns are clearly seen alongside the walls of the tube and around the ejected "bubble".

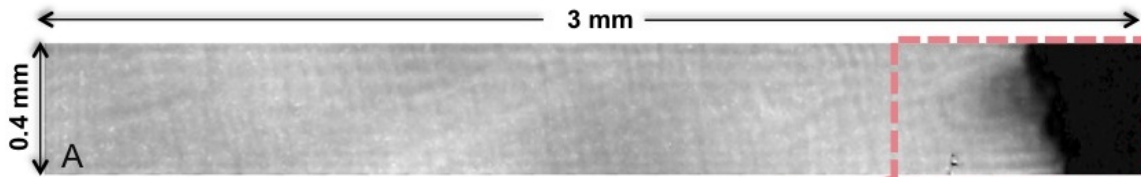

Fig.5- 3D representations and false color image of the transmitted XRL light from [1]. Only a 0.5 x 0.4 mm zone of the radiography is displayed, cleaned from the dust black mark. Undersampling by a factor of 2 has been used in the imageJ3D software. The attenuation ridge (the black zone in the false color image) probably correspond to the overdense shock peak. The attenuation gradient is clearly seen.

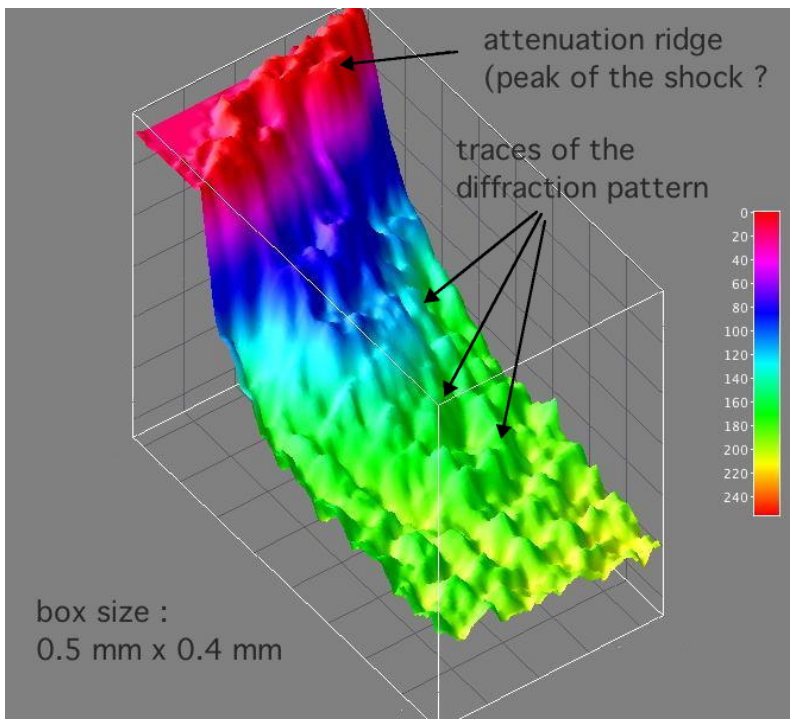

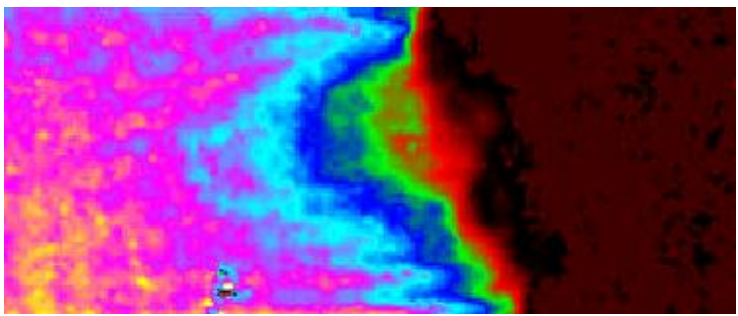

4- XRL image of a fired target, from [5], p. 33. The existence of a bubble diffraction pattern and of some crater at the back of the pusher appear to be reproducible features. Diffraction along the side of the tube and background variations are also reproducibly observed.

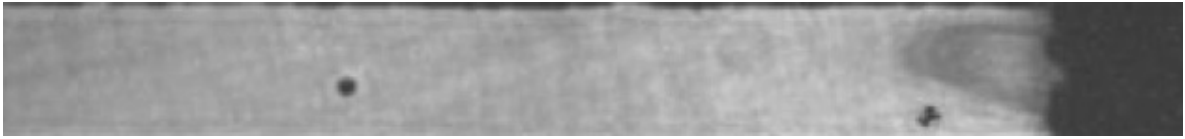

Fig.6:

3D and half-tone representations of pixel intensity from Ref.[5] (from 1 mm to 0.25 mm of the image right edge). Ejected matter coming from a back surface crater and creating the plowing jet-like structure is clearly seen. However discrimination between diffraction pattern and probe beam attenuation may be difficult. The dark spot is probably a dust grain. Undersampling by a factor of 2 has been used in the imageJ3D software.

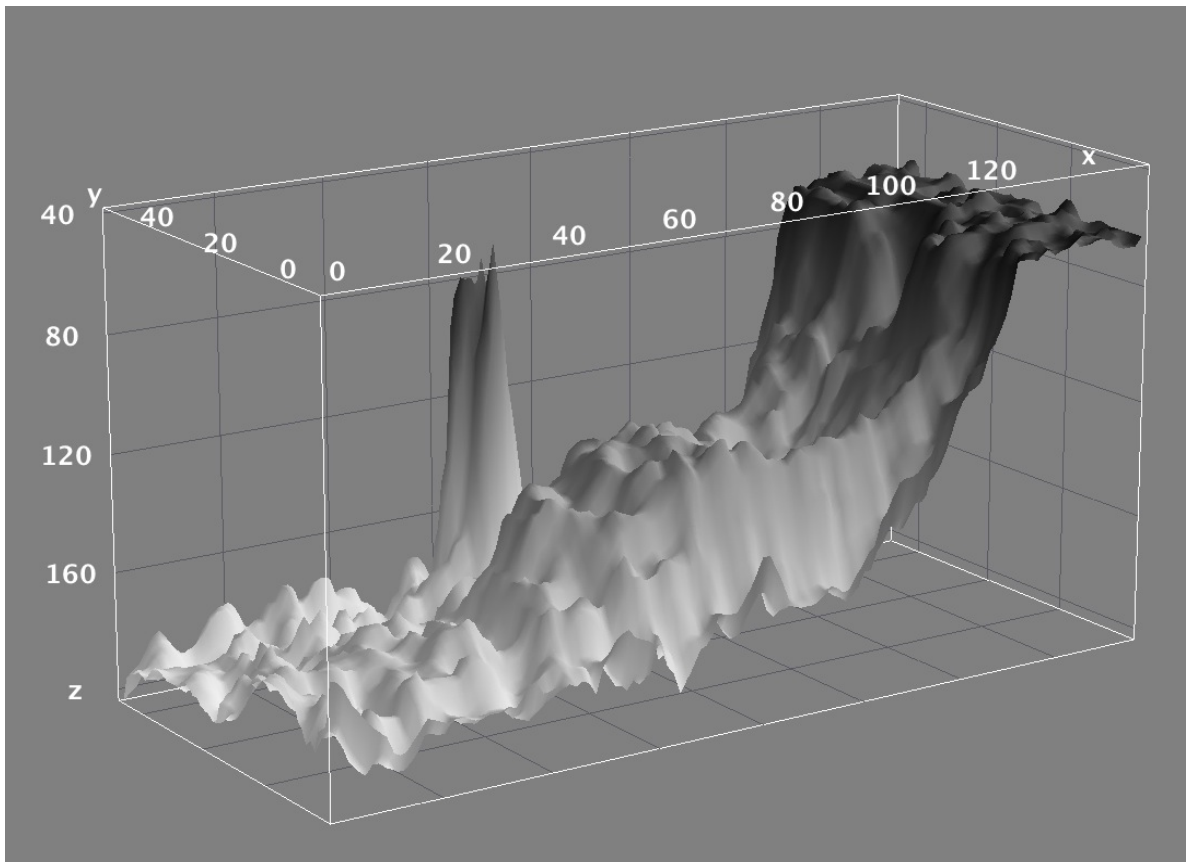

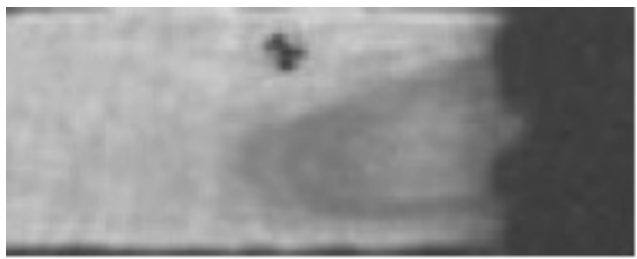

# MORE ANALYSIS

Fig.2 of SF2A 2013: Laser Experiment to study Radiative Shocks relevant to Astrophysics
U.Chaulagain, C.Stehlé et al.

- position of read and black lines are arbitrary (change the gray scale and …)
- what is labelled "precursor" is actually a diffraction pattern

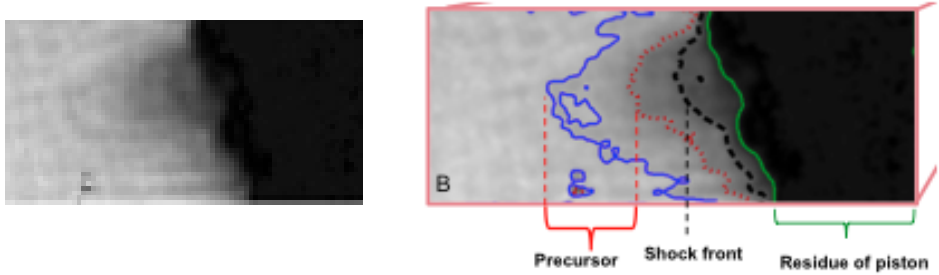

diffraction patterns :

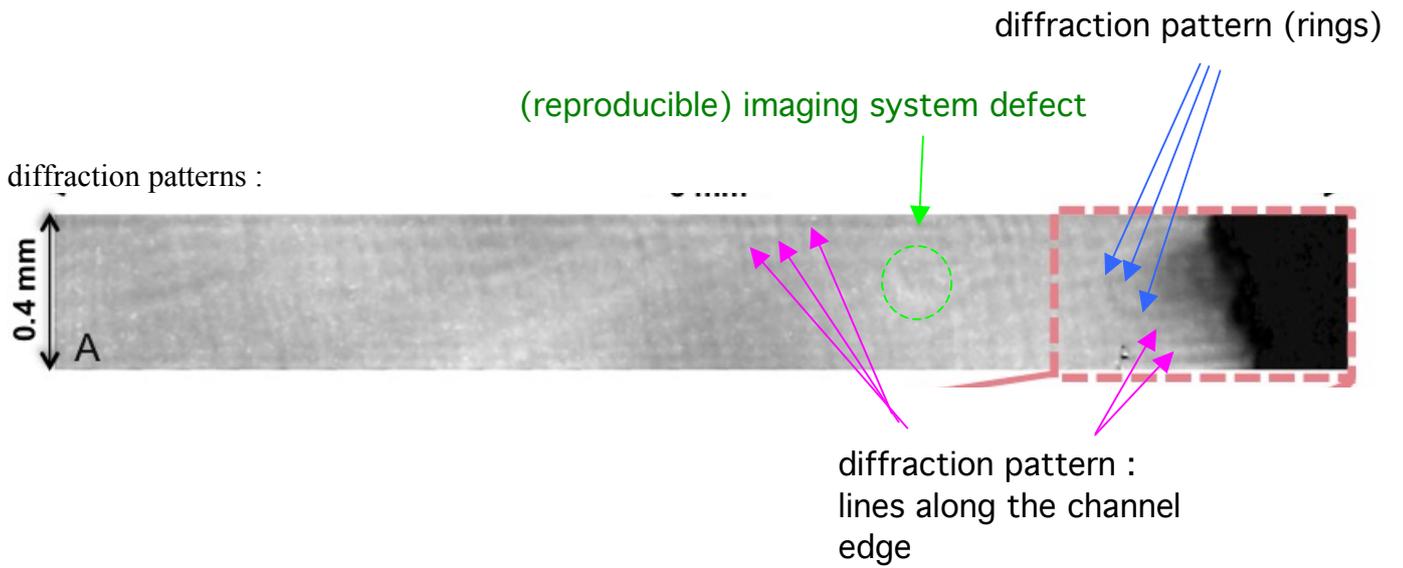

density gradient : 3D view of pixel intensity (with a 20 μm smoothing)

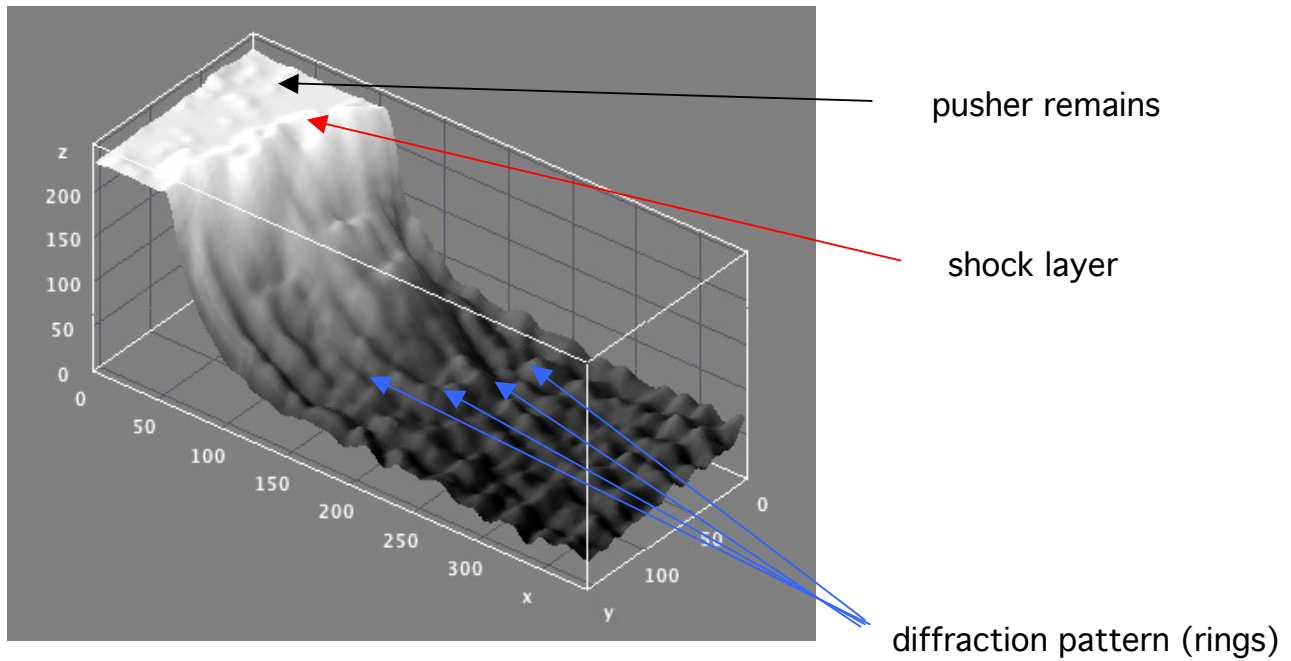

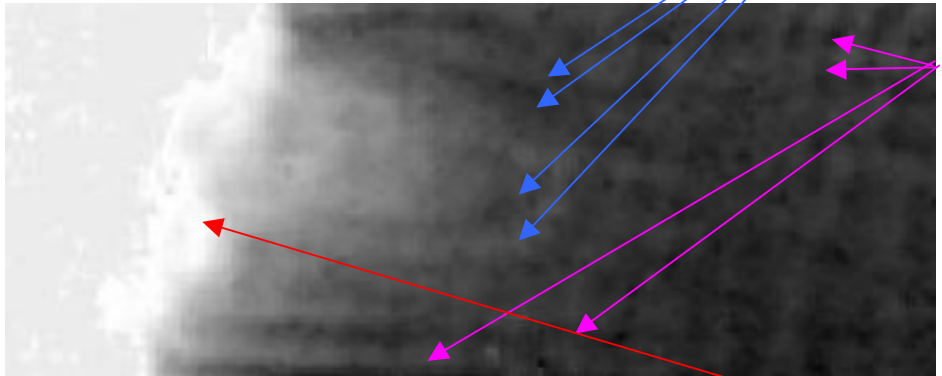
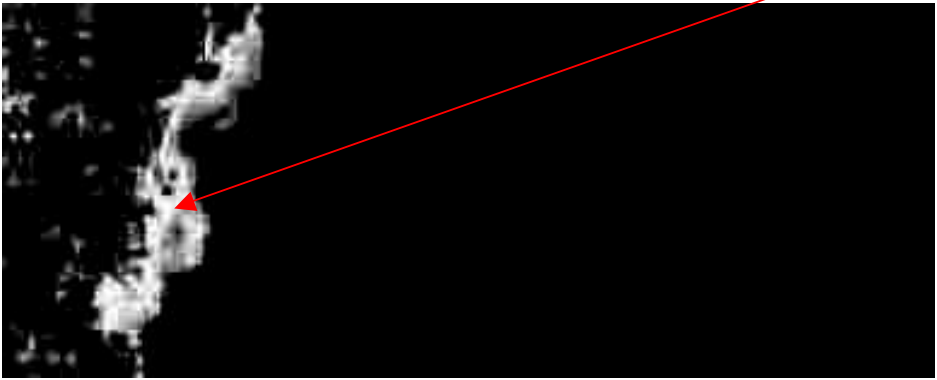

from Chaulagain et al (Kathmandu 2013):
expelled matter from crater at backside of the pusher seems to produce a bow wave (or a snow-plough effect) 1/4 mm away. The crater may be related to some laser focal spot properties.

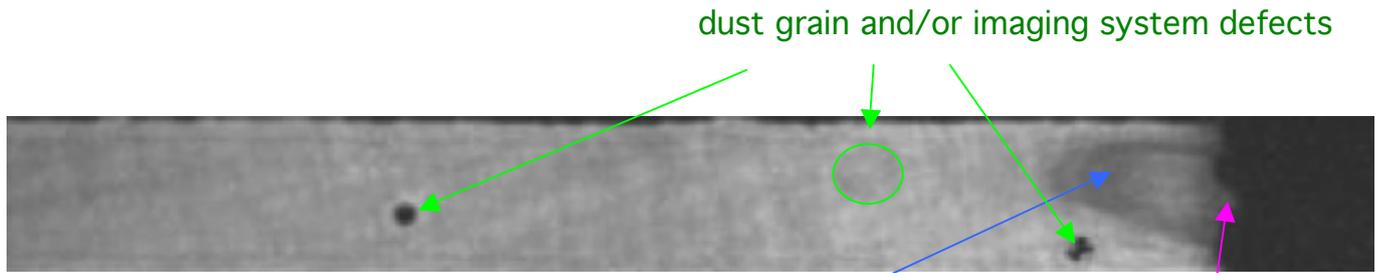

dust grain and/or imaging system defects

"bow wave" or snow-plough

crater

3D plot of pixel intensity (box size, approx 1 mm) :

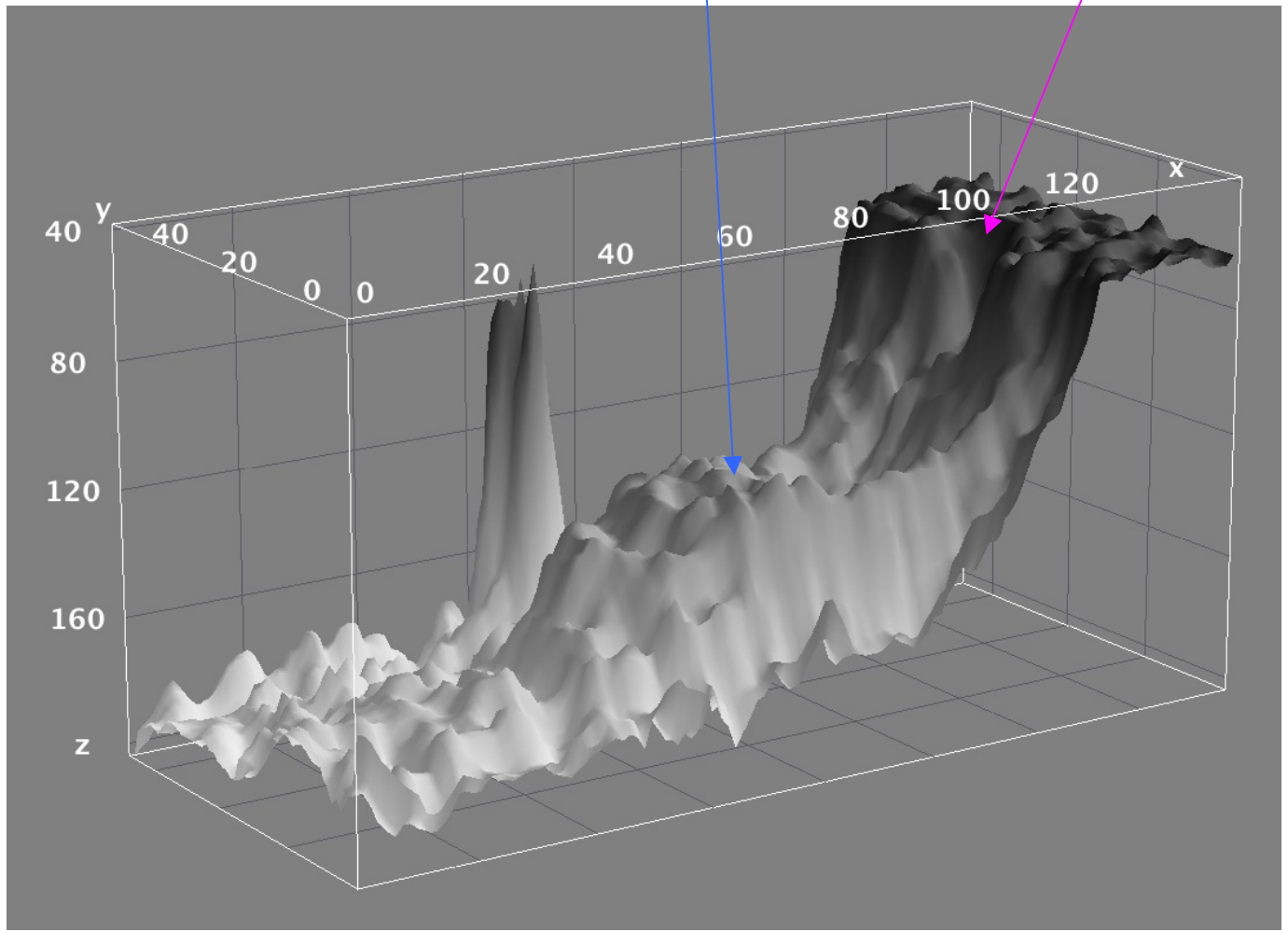